\documentclass{PoS}

\title{Long term radio variability of AGN}

\ShortTitle{Long term radio variability of AGN}

\author{\speaker{Talvikki Hovatta},$^1$ Merja Tornikoski,$^1$ Harry J. Lehto,$^{2,3}$ Elina Nieppola,$^1$ Esko Valtaoja,$^{2,3}$ Markku Lainela,$^2$ Ilona Torniainen,$^1$ Anne L\"ahteenm\"aki$^1$ Margo F. Aller$^4$ and Hugh D. Aller$^4$\\
\llap{$^1$}Mets\"ahovi Radio Observatory, TKK, Helsinki University of Technology\\
Mets\"ahovintie 114, 02540 Kylm\"al\"a, Finland\\
\llap{$^2$}Tuorla Observatory, University of Turku\\
V\"ais\"al\"antie 20, 21500 Piikki\"o, Finland\\
\llap{$^3$}Department of Physics, University of Turku\\
20140 University of Turku, Finland\\
\llap{$^4$}Department of Astronomy, University of Michigan\\
Ann Arbor, MI 48109, USA\\
E-mail: \email{tho@kurp.hut.fi}}

\abstract{A large number of AGN have been monitored for nearly 30 years at 
22, 37 and 87 GHz in Mets\"ahovi Radio Observatory. 
These data were combined
with lower frequency 4.8, 8.0 and 14.5 GHz data from the University of
Michigan Radio Astronomy Observatory, higher frequency data at 90 and 230 GHz 
from
SEST, and supplementary higher frequency data from the literature
to study the long-term variability of a large sample of AGN. Both the
characteristics of individual flares from visual inspection and
statistically-determined variability timescales as a function of
frequency and optical class type
were determined. Based on past behaviour, predictions of sources expected to
exhibit large flares
in 2008--2009 appropriate for study by GLAST and other instruments are made.
The need for long-term data for properly understanding source behaviour is
emphasised.}

\FullConference{Workshop on Blazar Variability across the Electromagnetic Spectrum\\
                 April 22-25, 2008\\
                 Palaiseau, France}

\begin{document}

\section{Introduction}
Active galactic nuclei (AGN) are variable across the whole electromagnetic 
spectrum. The radio regime is of special interest because we can directly 
observe the synchrotron radiation from the jet. The total flux density 
variations and flares seen in the flux curves are usually explained 
with shock-in-jet 
models where a disturbance creates a shock moving down in the jet, 
and as the shock develops we see a flare evolving from higher submm- and
mm wavelengths towards lower radio frequencies \cite{marscher85, hughes85}.

We have used a sample of 90 AGN to study their long-term 
variability timescales \cite{hovatta07, hovatta08b} and flare 
characteristics \cite{hovatta08a, nieppola08}. Our extensive 
database enables us to study the correspondence between the shock 
model and the observations, and also the statistical differences between 
the different AGN types (27 high polarisation quasars (HPQs), 33 low 
polarisation quasars (LPQs), 25 BL Lacertae objects (BLOs), and 5 radio 
galaxies (GALs)). Our sample 
includes bright sources that have flux density at least 1 Jy in the 
active state. 

\section{Variability timescales}
We used four statistical methods to study variability 
timescales in a sample of 80 AGN at 7 frequency bands between 4.8--230 GHz. 
We used the Structure Function (SF), 
the Discrete Correlation Function (DCF), the Lomb-Scargle periodogram, 
and wavelets. In the wavelet analysis we used only frequencies 22, 
37 and 90 GHz. In addition to analysing the timescales, we studied the 
properties and the differences of the methods. From the Fourier-based methods 
\cite{hovatta07}, the SF gives a timescale 
related to short variations such as the rise and the decay times 
of flares. The DCF and the Lomb-Scargle periodogram give the time 
between flares, and therefore indicates how often 
shocks are formed in the jet. We found that especially the Lomb-Scargle 
periodogram easily produces spurious spikes. The DCF is somewhat better but 
still suffers from treating a flux 
curve as one entity, a property of Fourier-based methods. 
Compared to Fourier-based methods, wavelets preserve 
the locality of the timescale i.e. show when and for how long it has been 
present in the flux curve. Another good property of the locality is that 
a change in a few 
points only affects the analysis locally. We lose, however, in the resolution 
of detecting the timescale, but when 
studying quasi-periodicities the accuracy is adequate. 
We also found that the DCF and the periodogram give very similar results 
as wavelets, but with wavelets it is possible to detect if the timescale 
is persistent or only short-lived.

\begin{figure}
\centering
\includegraphics[height=12cm]{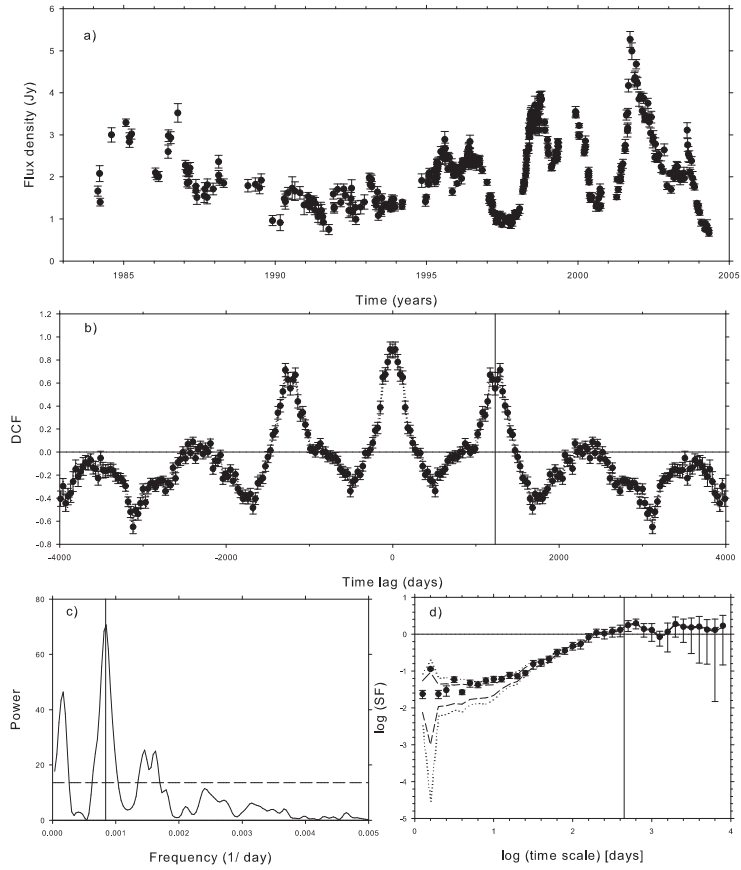}
\caption{Analyses of the quasar 4C 29.45 (1156+295) at 22 GHz. a) Flux density curve b) the DCF c) the Lomb-Scargle periodogram d) the SF. Timescales obtained with each method are marked by vertical lines. The periodogram timescale is 3.29 years which is 0.2 years shorter than the DCF timescale of 3.49 years. The SF gives a timescale of 1.21 years. \cite{hovatta07}}
\label{fig1}
\end{figure}

\begin{figure}
\centering
\includegraphics[angle=-90, width=.6\textwidth]{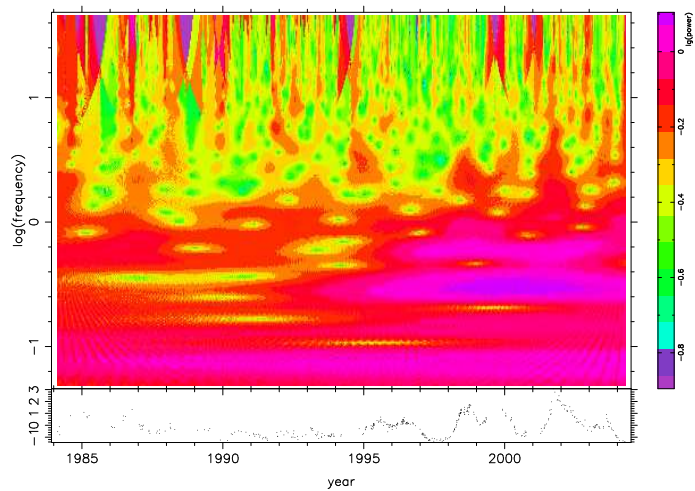}
\caption{The wavelet analysis of 4C29.45 (1156+295) at 22 GHz. The timescale 3.4 is obtained from the plot at $\log(10^{0.53})$ years. It is only present in the latter half of the flux curve starting at $\sim$ 1995. \cite{hovatta08b}}
\label{fig2}
\end{figure}

As an example, in Fig.\,\ref{fig1} the SF, the DCF and the 
Lomb-Scargle periodogram 
results at 22 GHz
are shown for the quasar 4C 29.45. The DCF in Fig.\,\ref{fig1}b gives a 
timescale of 3.49 years which is very close to the timescale 3.29 years 
obtained with the Lomb-Scargle periodogram (Fig.\,\ref{fig1}c). Indeed, when 
examining the flux curve in Fig.\,\ref{fig1}a it is possible to see flares 
with 3 years in between them. However, the wavelet analysis of the source in 
Fig.\,\ref{fig2} shows how the timescale 
(3.4 years from 
the wavelet plot) is present only in the latter half of the flux curve.
Thus by using only the Fourier-based methods, one could have claimed the 
source to be periodic but, instead, the behaviour of the source has changed 
over the monitoring period. Therefore we consider it appropriate to use 
wavelets when quasi-periodicities in AGN are studied. We also noticed 
that in many sources the timescales change slowly, get weaker in power, 
or disappear over long periods. Multiple timescales are common in these 
sources and are detected both in wavelet and in Fourier-based methods.

None of the sources in our sample showed strict periodicity in our analyses.
Flares are seen in these sources on average every 4 years at 22 and 37 GHz 
and the rise and the decay times of flares are between 1 to 2 years. 
This is also 
seen in Fig.\,\ref{fig1}d, where a SF timescale of 1.2 years is found for 
the source 4C 29.45. The average or the median timescales 
at three of the frequency bands are shown in Table \ref{table1}.  
We also studied the differences between HPQs, LPQs, 
and BLOs. The differences in their observed timescales were insignificant, 
and in all of them flares are observed, on average, in the same timescales.
When we studied the intrinsic redshift-corrected timescales we found  
indications that BLOs have longer timescales than quasars. 
Quasars seem to have flares every 2 years while in BLOs they happen 
intrinsically every 3 to 4 years. At 37\,GHz, 
the difference between BLOs and quasars in the DCF timescales was significant 
according to Kruskal-Wallis analysis, and in the wavelet timescales the 
BLOs differed significantly from the LPQs and there were indications that 
also the BLOs and the HPQs differ from each other. These results could 
imply that shocks are developed less frequently in BLOs than in quasars. 
This is contradictory to results from hydrodynamical 
simulations \cite{hughes02}, which show 
that shocks are formed more easily in slower jet flows which are often 
associated with BLOs. This will be studied further in a 
forthcoming paper \cite{hovatta08c}, where we will calculate the 
Lorentz factors for a large sample of sources from different classes.

\begin{table}
\centering
\begin{tabular}{lcccc}
\hline
\hline
Timescale    &   14.5 GHz  & 37 GHz  & 37 GHz z-corrected & 90 GHz\\
\hline
L-S periodogram (ave)  & 9.2  & 6.3  & 4.3  & - \\
DCF (ave) & 6.7 & 4.2 & 2.6   &  3.1 \\
Wavelet (ave) & - & 4.4 & 2.8  & 2.9 \\
SF (median) & 3.2 & 1.4 & 0.7 & 1.1 \\
duration (median) & 2.8 & 1.4 & - & 2.3 \\
\hline
\end{tabular}
\caption{Average or median timescales (in years) obtained with different methods at 3 different frequency bands. At 37 GHz also the redshift-corrected values are shown.}
\label{table1}
\end{table}

Even though none of the sources in our sample are strictly periodic, many of 
them show episodes of quasiperiodic behaviour. Therefore we were interested 
to see whether we could ``predict'' upcoming active states in these sources. 
It would be useful to know which sources are in an active state when planning 
multiwavelength observations. In particular, we were interested in sources 
which would be likely to flare during the early phases of the 
GLAST satellite, in the time period 2008--2009. Our earlier work 
(e.g. \cite{tornikoski02, lahteenmaki03}) has shown that strong 
gamma-ray activity is connected to a growing flare in the radio regime, 
and therefore sources that are active in the radio are good candidates to be 
detected by GLAST. 

Thus, based purely on our statistical wavelet analysis \cite{hovatta08b} 
combined with analyses of \cite{hovatta07} and visual inspection of 
our flux curves at 37 GHz until the end of 2007, we came up with a list 
of six potential sources. These sources show quasiperiodic behaviour in 
the wavelet analysis and in at least one Fourier-based method with 
such a timescale that a large flare could occur in 
2008--2009. Sources exhibiting almost continuous and complex variability, 
such as OJ 287 and BL Lac were excluded from the study. The six sources 
include one GAL type object (0007+106), two HPQs (0234+285, 1156+295), 
two LPQs (0333+321, 2145+067) and one BLO (1749+096). Out of these six 
sources, two (0007+106 and 1749+096) have redshifts less 
than 0.5 and could be potential sources to be detected at TeV 
energies as well. 

\section{Flare characteristics}
We have studied the flare characteristics of 55 sources with 159 
well-monitored flares at 8 frequency bands between 4.8--230 GHz. We calculated 
several parameters for each flare, e.g. the amplitude, the duration, and 
the time delays between the frequency bands. We were not able to calculate 
all the parameters for all the flares at every frequency band because 
especially the higher frequency data were often sparsely sampled.
The median peak flux density in the sample was found to be 4.5 Jy at 37 GHz.
The range of the peak flux densities in the sample was very broad, 
from 0.7 Jy to 57 Jy, showing that the variability behaviour in these sources 
is very heterogeneous, and studies concentrating on a single source are 
inadequate in describing the typical behaviour of AGN. This is also seen in 
the flare durations which we found to range between 0.3 and 
13.2 years. The median duration at 37 GHz is 2.5 years. The long duration of 
flares in the radio regime also show that long-term monitoring is 
essential in understanding the behaviour of AGN. Short multiwavelength
campaigns lasting from a couple of days to a few weeks are not sufficient 
in capturing a complete radio flare.

\begin{figure}
\centering
\includegraphics[width=.6\textwidth]{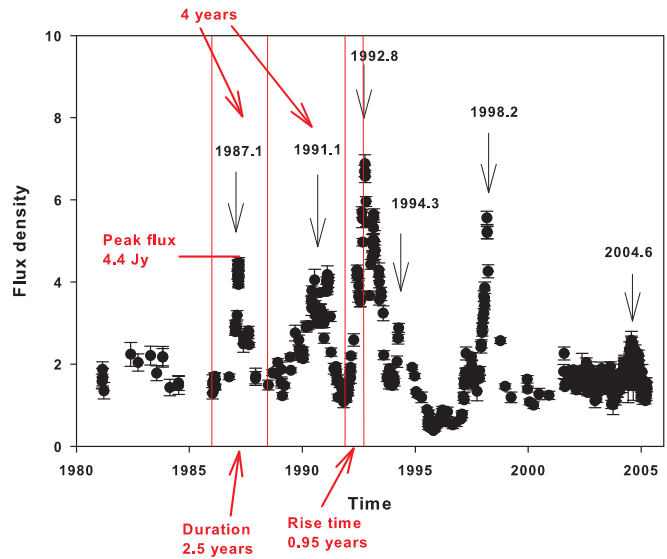}
\caption{Flux curve of an example source 0235+164 at 37 GHz.}
\label{fig3}
\end{figure}

An example of an ``average'' blazar is in Fig. \ref{fig3}, where 
the flux curve of the BLO source 0235+164 at 37 GHz is shown. In the flux 
curve of this source we can identify the median or average values of 
all the parameters from the different analyses. 
The flare peaking at 1987 has a peak flux density of 4.4 Jy, and also the 
median flare duration of 2.5 years is seen
in the same flare. We can also find the average time interval 4 years between 
the flares in 1987 and 1991. The rise time 0.95 years of the 1993 flare 
also agrees with the results of the SF analysis.
  
We compared the duration of the flares with the relative peak flux densities, 
and found that there is only a slight positive correlation of 0.15 between 
the values. We can see almost as high peak flux densities in flares 
lasting for 2 years as for flares lasting for 13 years.
The correlation disappears altogether when we correct the peak luminosity 
and the duration for redshift and Doppler boosting (Doppler boosting factors 
taken from \cite{hovatta08c}). This indicates that 
the energy release in a flare does not increase with the duration of 
the flares. 

Our sample includes sources from various AGN classes and 
are mostly blazars. The BL Lacertae objects and their timescales 
are studied in more detail in \cite{nieppola08}. In addition 
to the flare parameters, we studied the correspondence between the 
observations and the shock model by Marscher \& Gear \cite{marscher85} and its 
generalisation by Valtaoja et al. \cite{valtaoja92}. 
We found that the observed peak fluxes and the time delays between the 
flares adhere quite well to the predictions of the shock model, even 
though there is large scatter in the data. We note that our definition of 
a flare is based on visual inspection and therefore many of the events 
classified as flares in our analysis may include multiple shocks, increasing 
the scatter. In our analysis we found no 
major differences between the quasars and the radio-bright BLOs.

\section{Conclusions}
We studied the long-term radio variability of a sample of 90 sources 
using statistical timescale analysis methods and visual inspection of 
the flare parameters. Our main results are the following:

\begin{itemize}

\item{Fourier-based methods the DCF and the periodogram give very similar results as wavelets, but wavelets should be used when quasi-periodities are studied because they give information on the locality of the timescale. With wavelets it is possible to see if a timescale is long-lasting or just a short transient phenomenon in the flux curve. DCF or periodograms can then be used to verify the timescale more accurately.} 

\item{Variability behaviour is complex and no clear periodicities could be 
found at radio frequencies. Episodes of quasi-periodic behaviour are common, and therefore false periodicities may be found if the temporal coverage is inadequate.}

\item{Flares are seen, on average, every 4 years in all the source types at 37\,GHz but when intrinsic redshift-corrected timescales are studied, the quasars have shorter timescales of 2 years compared to the 3-4 years of BLOs. This could indicate that shocks are produced less frequently in BLOs than in quasars.}

\item{Median duration of a flare is 2.5 years at 22 and 37\,GHz, but the range in durations is between 0.3 and 13.2 years. When comparing the duration with intrinsic redshift- and Doppler-corrected peak luminosities, we found that the energy release in a flare does not increase with the duration of the flare.}

\item{Flares adhere quite well to the predictions of the shock model but the scatter in the data, due to poor sampling and complicated structure of the flares, is still large.}

\item{By combining the median duration of flares, 2.5 years, with the average 
time between the flares, 4 years, we see that multifrequency campaigns 
should last for 5-7 years in order to catch the source in both its 
highest and lowest activity states.}

\item{Long-term monitoring is essential in understanding the true behaviour 
of these sources at radio frequencies.}
\end{itemize}


\begin{thebibliography}{99}
\bibitem{marscher85}A. P. Marscher and W. K. Gear, \emph{Models for high-frequency radio outbursts in extragalactic sources, with application to the early 1983 millimeter-to-infrared flare of 3C 273}, \emph{ApJ} {\bf 298} (1985) 114M
\bibitem{hughes85}P. A. Hughes, H. D. Aller and M F. Aller, \emph{Polarized Radio Outbursts in Bl-Lacertae - Part Two - the Flux and Polarization of a Piston-Driven Shock}, \emph{ApJ} {\bf 298} (1985) 301H
\bibitem{hovatta07}T. Hovatta, M. Tornikoski, M. Lainela, H. J. Lehto, E. Valtaoja, I. Torniainen, M. F. Aller, and H. D. Aller, \emph{Statistical analyses of long-term variability of AGN at high radio frequencies}, \emph{A\&A} {\bf 469} (2007) 899H
\bibitem{hovatta08b}T. Hovatta, H. J. Lehto and M. Tornikoski, \emph{Wavelet analysis of a large sample of AGN at high radio frequencies}, \emph{A\&A} (2008) in press
\bibitem{hovatta08a}T. Hovatta, E. Nieppola, M. Tornikoski, E. Valtaoja, M. F. Aller and H. D. Aller, \emph{Long-term radio variability of AGN: flare characteristics}, \emph{A\&A} {\bf 485} (2008) 51H
\bibitem{nieppola08}E. Nieppola, T. Hovatta, M. Tornikoski, E. Valtaoja, M. F. Aller and H. D. Aller, \emph{Long-term variability of radio-bright BL Lacertae objects}, (2008) in preparation
\bibitem{hughes02}P. A. Hughes, M. A. Miller and G. C. Duncan, \emph{Three-dimensional Hydrodynamic Simulations of Relativistic Extragalactic Jets}, \emph{ApJ}  {\bf 572} (2002) 713H
\bibitem{hovatta08c}T. Hovatta et al. in preparation
\bibitem{tornikoski02}M. Tornikoski, A. L\"ahteenm\"aki, M. Lainela, and E. Valtaoja, \emph{Possible Identifications for Southern EGRET Sources}, \emph{ApJ} {\bf 579} (2002) 136T
\bibitem{lahteenmaki03}A. L\"ahteenm\"aki and E. Valtaoja, \emph{Testing of Inverse Compton Models for Active Galactic Nuclei with Gamma-Ray and Radio Observations}, \emph{ApJ} {\bf 590} (2003) 95L 
\bibitem{valtaoja92}E. Valtaoja, H. Ter\"asranta, S. Urpo, N. S. Nesterov, M. Lainela and M. Valtonen, \emph{Five Years Monitoring of Extragalactic Radio Sources - Part Three - Generalized Shock Models and the Dependence of Variability on Frequency}, \emph{A\&A} {\bf 254} (1992) 71V

\end{thebibliography}
\end{document}